# Cation-like Doppler shifts from a neutral molecule in an electrical discharge.

Hervé Herbin, Robert Farrenq, Guy Guelachvili, Nathalie Picqué

Laboratoire de Photophysique Moléculaire, Unité Propre du CNRS,
Bâtiment 350, Université de Paris-Sud, 91405 Orsay Cedex, France.

Corresponding author:
Dr. Nathalie Picqué,
Laboratoire de Photophysique Moléculaire
Unité Propre du CNRS, Université de Paris-Sud, Bâtiment 350
91405 Orsay Cedex, France
Website: http://www.laser-fts.org
Phone nb: 33 1 69 15 66 49
Fax nb: 33 1 69 15 75 30
Email: nathalie.picque@ppm.u-psud.fr

## Abstract

Velocity-modulation Fourier transform emission spectra from a $N_2O$/He discharge plasma recorded between 1 and 5.5 µm are described. Surprisingly, they show Doppler-shifted lines for the $E\ ^2\Sigma^+$- $D\ ^2\Sigma^+$, $C\ ^2\Pi$- $A\ ^2\Sigma^+$, and $D\ ^2\Sigma^+$- $A\ ^2\Sigma^+$ Rydberg-Rydberg rovibronic transitions of the nitric oxide neutral molecule. These polarity-dependent Doppler-shifts are those of positively charged particles. Vibration-rotation lines of NO and transitions from other neutral molecules like $N_2$ are also present in the spectra with comparable intensities and remain unshifted. Experimental investigations and possible explanations are discussed.

## 1. Introduction

Selective detection spectroscopic techniques are attractive because they discriminate transient molecules against the more abundant precursor molecules. As a consequence they have been popular for a long time in laser absorption spectroscopy. Amongst them velocity modulation (VM) [1] has been since 1983 successfully implemented for the search and characterization of molecular ions. Selective approach has not been widely adopted for high-resolution work in Fourier Transform Spectroscopy (FTS). Double modulation FTS (DMFTS) [2,3] developed at Laboratoire de PhotoPhysique Moléculaire (Orsay, France) a few years ago, enable the simultaneous acquisition of selective and non-selective high-resolution emission or absorption interferograms. The resulting combined advantages of broad spectral coverage and accurate wavenumber determination have for instance proven to be useful [4] for Doppler shifts measurements leading to an investigation on the vibrational and rotational dependence of the average mobility of $ArH^+$ in the positive column of an Ar/He glow discharge.

In this letter, emission spectra from an $N_2O$/He discharge plasma recorded between 1800 and 10 000 $cm^{-1}$ with DMFTS using VM selection are presented. Surprisingly, the selective spectra show strong first-derivative-type lines of Rydberg-Rydberg rovibronic transitions of the nitric oxide neutral molecule. This results from Doppler shifts that would be encountered by positively charged particles. Polarity-dependent lineshifts from neutrals in glow discharges have already been seldomly reported. They are restricted to Doppler shifts in the light molecules $H_2$ [5,6] or He [7,8] induced by momentum transferred from colliding electrons. Consequently the neutral behave like anions. To our knowledge, this experiment presents, with NO, the first high resolution spectroscopic detection of a cation-like drift-velocity of a neutral molecule. Detailed description of the experimental investigations is given and tentative explanations are discussed.

## 2. Experiment

Experimental observations described in this paper were initiated by unexpected results while running preliminary tests for another experiment. The VMFT spectra of a $N_2O$/He discharge exhibited strong first-derivative-type profiles belonging to the $E\ ^2\Sigma^+$- $D\ ^2\Sigma^+$, $C\ ^2\Pi$- $A\ ^2\Sigma^+$, and $D\ ^2\Sigma^+$- $A\ ^2\Sigma^+$ Rydberg-Rydberg rovibronic transitions of NO. Various experimental conditions were then explored to investigate the phenomenon. They are presented hereafter .

The emission source is a discharge tube, identical to the one described in [9]. It has an inner diameter equal to 0.7 cm and the discharge zone length is 25 cm long. The gas $N_2O$/He (or $N_2O$ or NO/He) flows through the tube at a constant rate from a central inlet to two exit ports located at both ends. Gas pressures are not measured inside the discharge. As a consequence, their ratio is more accurate than their absolute values. The discharge is applied between two water-cooled stainless steel electrodes, one electrode being kept to the ground. Both *ac*- and *dc*- excitations of the plasma are used. The *ac* high voltage power-supply consists by two audio amplifiers, stepped up by two 50-kHz 3-dB bandwidth transformers. The emitting plasma is optically conjugated to the high-resolution stepping-mode interferometer of Laboratoire de Photophysique Moléculaire.

When the plasma is produced in an *ac* discharge, a selective VMFT spectrum is recorded simultaneously with the usual non-selective emission spectrum. A detailed description of the acquisition procedure is given in [2]. Briefly, path difference modulation is practiced as usual with stepping mode FTS and related synchronous detection leads to the classical non-selective interferogram. Additionally, velocity modulation provides the selective detection of ionic species. An alternative electric field inside a discharge changes the net drift-velocity of the ions at the discharge excitation current frequency. This results in synchronous Doppler shifts of the frequency of their transitions. An appropriate lock-in detection keeps only the field-sensitive transitions.

The experimental conditions of the spectra are summarized in Table 1. In the 5500-10000 $cm^{-1}$ region, when using $N_2O$/He as precursors, the DM spectrum (nb.3534 in Table 1) illustrated on Figs. 1, 2 and 3 is observed. As shown on Fig.1, the nonselective component reports $N_2$ lines from the $B\ ^3\Pi_g$ - $A\ ^3\Sigma^+_u$ ($\Delta\upsilon$ = -2, -1 and 0) [10] as well as $B'\ ^3\Sigma^-_u$ - $B\ ^3\Pi_g$ (1-0) [11] transitions not indicated on the figure. NO lines from the $E\ ^2\Sigma^+$ - $D\ ^2\Sigma^+$, $C\ ^2\Pi$ - $A\ ^2\Sigma^+$, and $D\ ^2\Sigma^+$ - $A\ ^2\Sigma^+$ ($\Delta\upsilon$ =0) [12,13] Rydberg-Rydberg rovibronic transitions are also present. The selective component is only made of NO transitions. In spite of the advantage of a higher frequency modulation, NO transitions in the selective component are seen with smaller signal-to-noise ratio (SNR) respectively reduced by a factor of 20, 6 and 7 for the *E-D*, *C-A* and *D-A* transitions. This is due Doppler shifts smaller than half-widths of line profiles [2]. With an expanded scale the NO *C-A* transition is shown on Fig. 2 in order to better illustrate the selectivity efficiency. A small portion of the spectrum is again expanded on Fig. 3 where the selective component appears free from $N_2$ lines and the NO lineshapes have the expected appearance induced by the VM approach.

In order to check that instrumental artefacts were not at the origin of this unexpected VM detection of NO, additional spectra listed on Table 1 were recorded. Classical spectra 3538 and 3539 sequentially recorded with opposite *dc* excitations of the discharge confirmed real shifts for only NO lines together with their cation-like behavior. Figure 4 reports the individual lineshifts extracted from these two spectra for each of the NO transitions. Particular attention was also brought to the possible shift effects due to improper optical connection between the discharge tube and the interferometer. DM spectrum 3540 obtained with a reduced entrance Jacquinot-stop (J-stop) diameter (1.4 intead of 5.7 mm) still presents the NO lineshifts with the same reduction factor on the respective SNR in the selective and nonselective components.

Precursor gas composition influence was also examined. Discharge in $N_2O$ (spectrum nb.3551) produces similar results with a SNR about 5 times lower for the NO Rydberg transitions. SNR reduction factors from nonselective to selective components are still the same as in spectrum 3534 for NO Rydberg lines. Looking for a similar shift on the 1-0 vibration-rotation band of NO in the X $^2\Pi$ electronic state revealed unsuccessful. Spectrum (nb. 3554) in the 1800-6500 $cm^{-1}$ range with $N_2O$/He as gas precursors displays easily observed (1-0) NO and $N_2O$ vibration-rotation bands, which both appear only in the nonselective component. When using NO/He as a precursor (spectrum nb.3560), the nonselective component looks approximately the same as with $N_2O$ on Fig.1, but no spectral signature is observed on the selective component.

## 3. Tentative interpretation.

In summary, with $N_2O$/He as precursors, the line positions of the Rydberg-Rydberg NO transitions encounter an apparent systematic polarity-dependent wavenumber shift of the order of 2 $10^{-3}$ $cm^{-1}$, while $N_2$ lines with similar intensity remain unshifted. Moreover, the observed shifts of NO in Rydberg states correspond to those of a cation. Furthermore no ion transitions are detected. The presence of the NO Doppler shifts is gas precursor dependent. Shifts are only observed in the experiments made with $N_2O$ or $N_2O$ + He. However, with $N_2O$ +He no shift was measurable in the observed fundamental 1-0 NO vibration-rotation band.

Several qualitative conclusions can be settled from the experimental investigations summarized above. Since NO line shifts are discharge-polarity dependent, they are not related to shifts or splittings of the energy levels (for instance, Stark effect). In the complete set of spectra evoked above, no molecular ion transitions could be identified. $N_2^+$ and $NO^+$ transitions were specifically looked for within the explored spectral range (1800-10 000 $cm^{-1}$). $N_2O^+$ could not be probed. All its known transitions [14] lie in the near UV range, which is not currently accessible to our experimental facility. Since NO behaves like a positively charged particle, the observed Doppler shifts cannot be due to momentum transferred in the electron-molecule collisions, as already reported for $H_2$ [5,6] and He [7,8].

Two hypotheses: electron-$NO^+$ recombination and momentum transfer in positive ion-molecule collisions, may be examined with more details. An electron - $NO^+$ recombination could indeed be the origin of the observations: $NO^+$ drifts under electric field influence and recombines with an electron to form NO in Rydberg states. If $NO^+$ forms NO in a time short compared to the radiative lifetime of the ion, the absence of $NO^+$ transitions in the spectra is not contradictory. Observation of different Doppler shifts for the three vibronic transitions could be due to state-dependant rate constant. However, with this assumption, we did not find any satisfactory explanation to the lack of Doppler-shifted NO transitions when NO is used as a discharge precursor (while these intense NO Rydberg-Rydberg transitions are observed in the nonselective spectrum). The second hypothesis is that NO Doppler shifts could derive from momentum transfer in positive ion - molecule collisions. The involved positive ion might be $N_2O^+$. Furthermore, the interaction potential between an ion and a polar molecule is given [15] by: $V(r,\theta) = -\alpha\, q^2/2r^4 - \mu\, q \cos\theta/r^2$ , where $r$ is the ion-dipole distance, $q$ is the ion charge, $\alpha$ and $\mu$ are respectively the polarizability and dipole moment of the neutral molecule and $\theta$ is the angle between the dipole and the ion-dipole directions. Dipole moments and polarizability [16] of Rydberg states are usely larger than valence states. In particular, Ref. [17] and [18] respectively report $\mu$(NO, $X^2\Pi$ ($\upsilon$ =1))=0.141 D, $\mu$(NO, $D^2\Sigma^+$)=2.21 D. This should lead to more important collisional state dependant cross-sections for NO when it is in Rydberg states than in valence states.

With the above second assumption, a simple modelling of the lineshape has been implemented. Cationic molecules, drifting under electric field influence, collide elastically a given proportion of the NO molecules in Rydberg states. Both types of molecules are being considered as hard spheres. A classical momentum transfer calculation shows that, after collision, NO molecules acquire, following an equiprobable distribution, a velocity component along the electric field axis varying between 0 and $\upsilon$. As a consequence, in the nonselective component, a lineshape centered at wavenumber $\nu_0$ is considered to be the superposition of a non-shifted Doppler profile and of the integral on the velocity distribution of Lorentz profiles with Doppler shift varying between $+\delta_{max} \sim \nu_0\ \upsilon/c$ and $-\delta_{max}$. VM lineshapes are derived and discussed in [2,4,19,20]. In the present simple calculation, a lineshape in the selective component is made of the difference between the sum of the red-shifted Lorentz profiles and the sum of the blue-shifted Lorentz profiles. In practice the non-shifted Doppler profile was numerically determined from $N_2$ line shapes and twenty Lorentz profiles were used to represent the velocity distribution. Individual line fits of a few NO *C-A* profiles in the selective and nonselective components were performed. An example is given on Figure 5. The fits provide a rough estimate of the proportion of the population of each state undergoing collisions and of the maximum Doppler shift. They are respectively of the order of 20 % and 8 $10^{-3}$ $cm^{-1}$ (corresponding to a drift-velocity of about 300 $m.s^{-1}$). Reciprocally, computed line profiles based on these parameters in *dc* excitation conditions enable to reproduce the order of magnitude of the observed Doppler shift from line position subtraction.

Acknowledgments
The referee is warmly acknowledged for bringing arguments discussing the electron - $NO^+$ recombination hypothesis.

| Spectrum Nb. | $N_2O$ pressure (Pa) | NO pressure (Pa) | He pressure (Pa) | *ac* / *dc* excitation | Free spectral range ($cm^{-1}$) | Non-apodized Resolution ( $10^{-3}$ $cm^{-1}$) | NO line shifts ? |
|---|---|---|---|---|---|---|---|
| 3534 | 350 | - | 350 | *ac* | 5 900 - 11 800 | 29 | Yes |
| 3538 | 350 | - | 350 | + *dc* | 5 900 - 11 800 | 29 | Yes |
| 3539 | 350 | - | 350 | - *dc* | 5 900 - 11 800 | 29 | Yes |
| 3540 (with small J-stop) | 350 | - | 350 | *ac* | 5 900 - 11 800 | 29 | Yes |
| 3551 | 400 | - | - | *ac* | 5 900 - 11 800 | 29 | Yes |
| 3554 | 350 | - | 350 | *ac* | 1 800 - 6 100 | 10 | No |
| 3560 | - | 350 | 250 | *ac* | 5 500 - 11 000 | 27 | No |

Table 1: Recording conditions. 3538 and 3539 are the only classical FT spectra. All other spectra are Double Modulation spectra.

When using *ac* excitation, typical discharge conditions are 950 V – 0.25 A, with 10 kHz squarewave modulation.

The interferometer is equipped with two InGaAs photodiodes when recording interferograms in the 5500 – 12000 $cm^{-1}$ region and with two liquid-nitrogen cooled InSb detectors (with an additional filter) in the 1800-6000 $cm^{-1}$.

Figure captions

Figure 1: Double modulation high resolution spectrum no 3534 (see Table 1). Surprisingly, NO Rydberg-Rydberg rovibronic transitions appear selectively in the velocity modulation component, which is supposed to retain only ionic species lines.

Figure 2: Expanded portion of Figure 1 showing the $C\ ^2\Pi$- $A\ ^2\Sigma^+$ (0-0) band of $^{14}N^{16}O$. The apodized resolution is 6 $10^{-2}$ $cm^{-1}$. The lack of transitions around 8280 $cm^{-1}$, made obvious in the selective component is due to a perturbation.

Figure 3: Illustration of the "selectivity": only NO lines appear in the selective component of the double modulation spectrum.

Figure 4: Individual line-shift of the $E\ ^2\Sigma^+$- $D\ ^2\Sigma^+$, $C\ ^2\Pi$- $A\ ^2\Sigma^+$, and $D\ ^2\Sigma^+$- $A\ ^2\Sigma^+$ transitions of NO. It is measured from the difference between line positions of two spectra numbered 3538 and 3539 in Table 1, recorded successively with opposite discharge polarities. The sign of these systematic shifts corresponds to those of a cation. Mean values of the shift are: 1 $10^{-3}$ $cm^{-1}$ for the *E-D* (0-0) band, 2 $10^{-3}$ $cm^{-1}$ for the *C-A* (0-0) band, 3 $10^{-3}$ $cm^{-1}$ for the *D-A* (0-0) band.

Figure 5: Line fit of the $R_{y1}(15.5)$ of the *C-A* (0-0) transition of NO. The selective and nonselective profiles have been normalized. The continuous lines represent the calculated profiles. Four other lines from the C-A (0-0) band have been fitted and provide a consistent estimate of the maximum Doppler shift, equal to 8 $10^{-3}$ $cm^{-1}$.

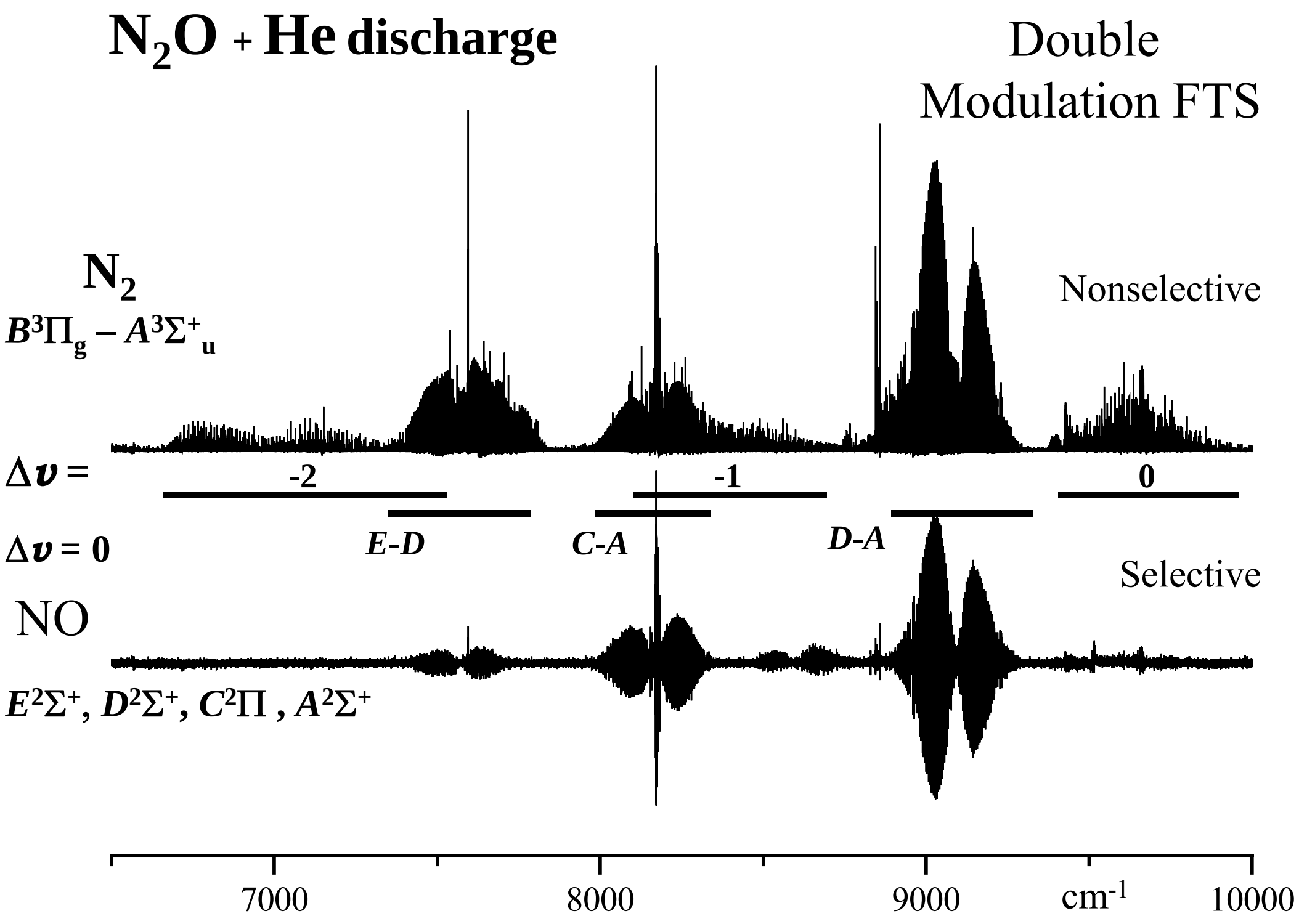

Figure 1

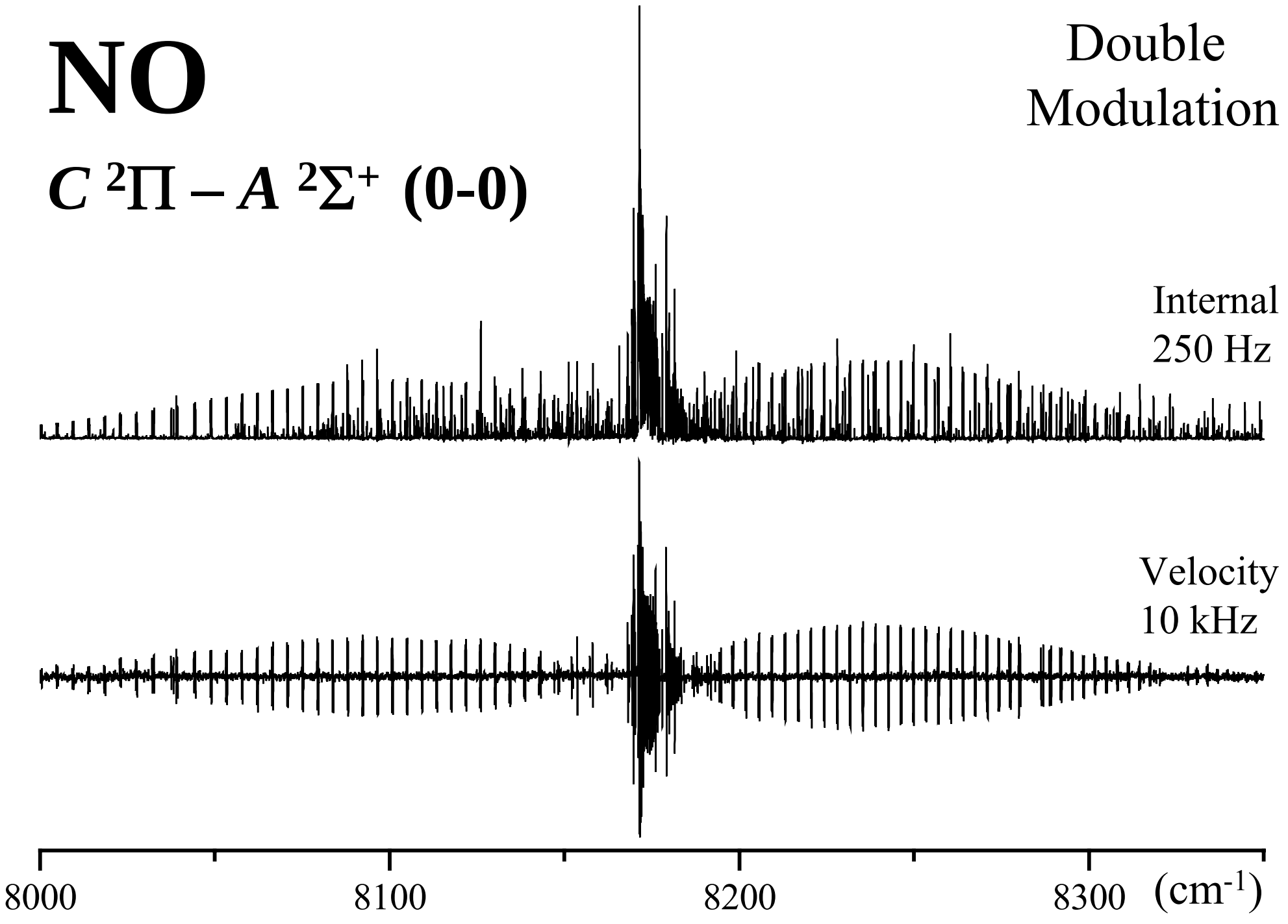

Figure 2

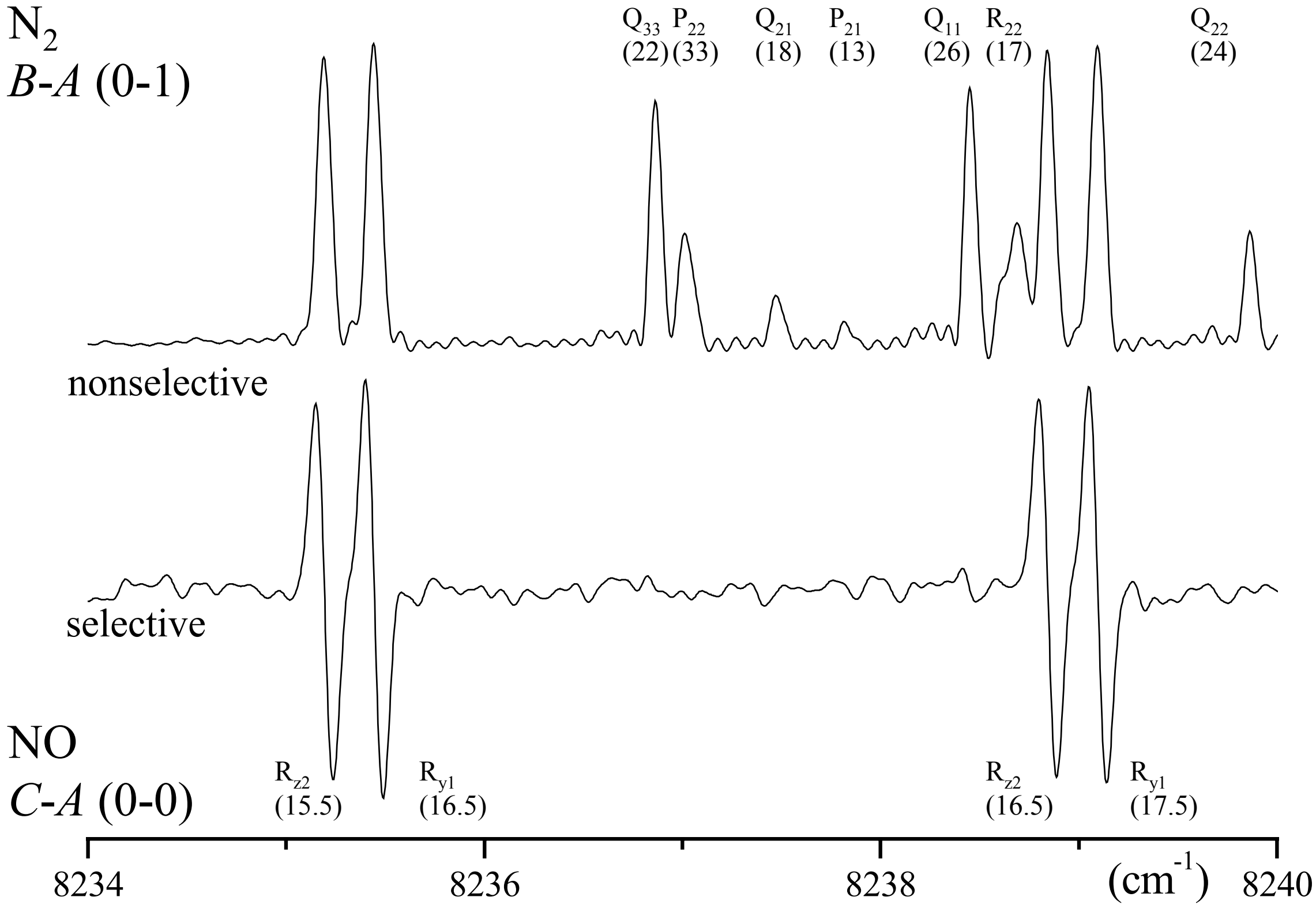

Figure 3

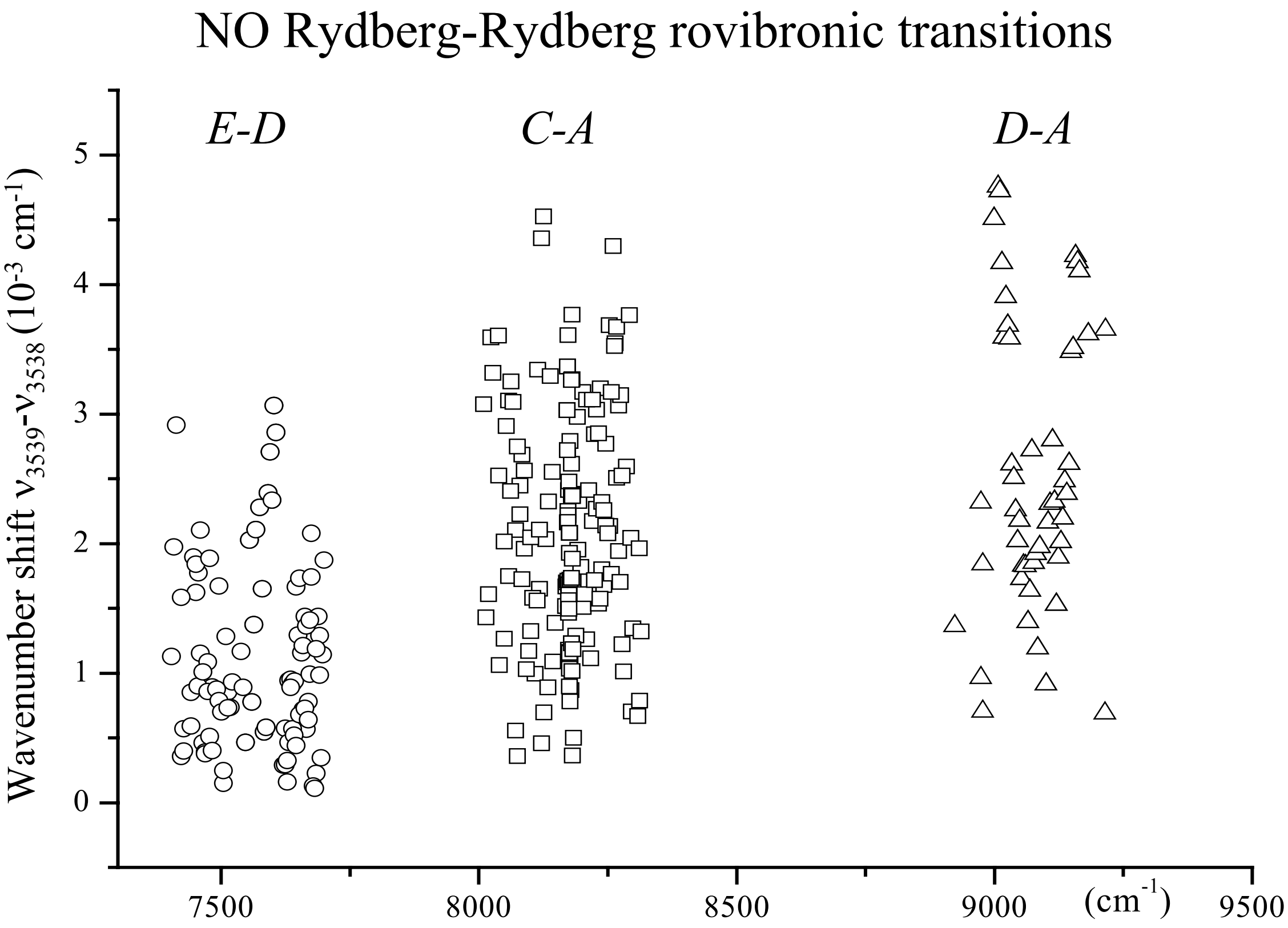

Figure 4

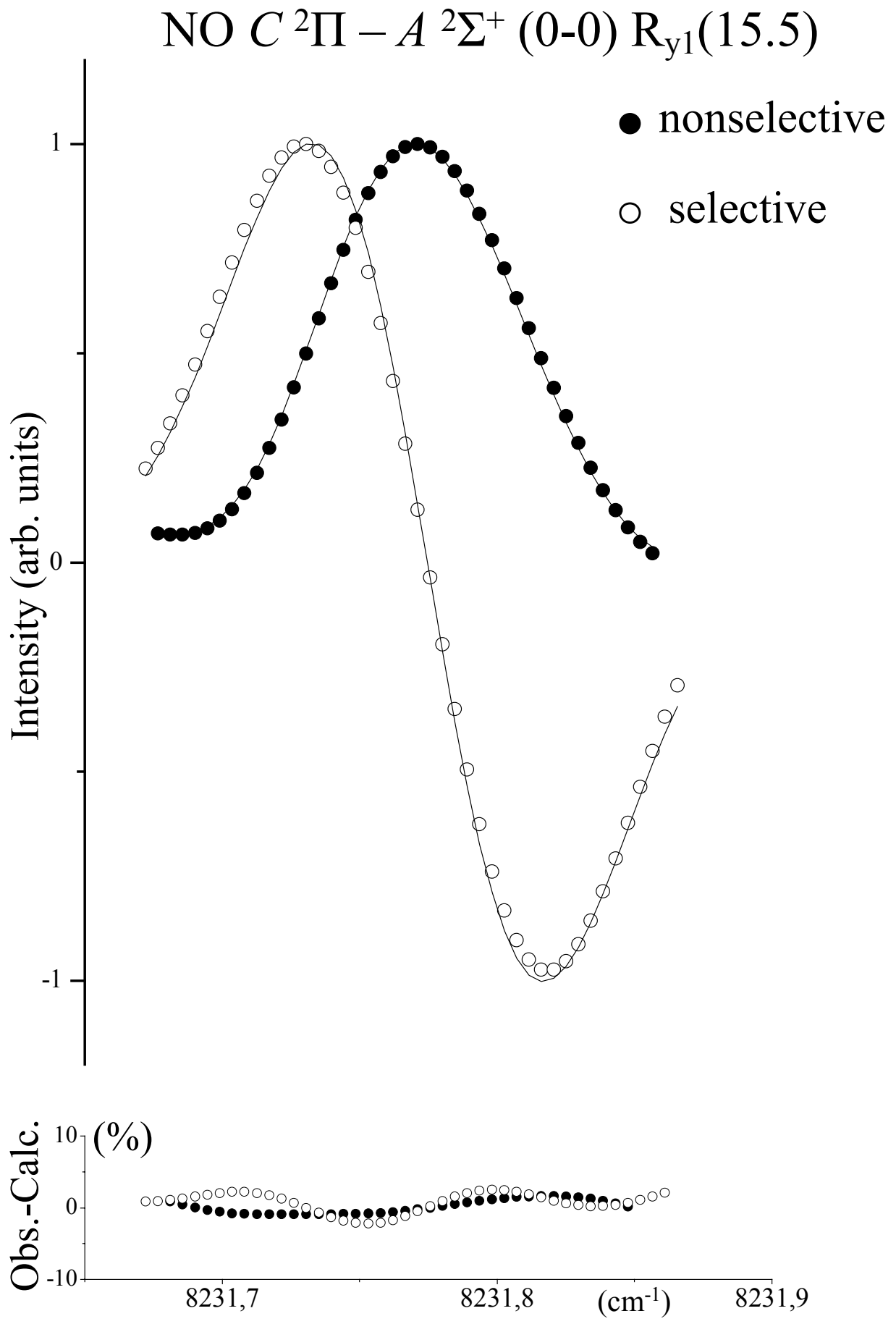
NO $C\ ^2\Pi - A\ ^2\Sigma^+$ (0-0) $R_{y1}$(15.5)
● nonselective
○ selective
Intensity (arb. units)
1
0
-1
Obs.-Calc.
10
(%)
0
-10
8231,7
8231,8
($cm^{-1}$)
8231,9

Figure 5